# Rosetta Alice Ultraviolet Spectrograph Flight Operations and Lessons Learned

Jon P. Pineau[1], Joel Wm. Parker[2], Andrew J. Steffl[2], Eric Schindhelm[2], Richard Medina[2], S. Alan Stern[2], M. Versteeg[3], and Emma M. Birath[2]

[1] Stellar Solutions, Inc., Palo Alto, CA, United States.
[2] Southwest Research Institute, Department of Space Studies, Boulder, CO, United States.
[3] Southwest Research Institute, 6220 Culebra Road, San Antonio, TX 78238, United States.

## Abstract

This paper explores the uniqueness of ESA Rosetta mission operations from the Alice instrument point of view, documents lessons learned, and suggests operations ideas for future missions. The Alice instrument mounted on the Rosetta orbiter is an imaging spectrograph optimized for cometary far-ultraviolet (FUV) spectroscopy with the scientific objectives of measuring properties of the escaping gas and dust, and studying the surface properties, including searching for exposed ices. We describe the operations processes during the comet encounter period, the many interfaces to contend with, the constraints that impacted Alice, and how the Alice science goals of measuring the cometary gas characteristics and their evolution were achieved. We provide details that are relevant to the use and interpretation of Alice data and published results. All these flight experiences and lessons learned will be useful for future cometary missions that include ultraviolet spectrographs in particular, and multi-instrument international payloads in general.

# 1 The Rosetta Mission and Comet 67P

At first sight of the comet 67P/Churyumov-Gerasimenko, it was obvious that the Rosetta mission would be different from any that came before. While this was expected to some extent as reflected in the mission design, the need for an even more complex concept of operations was realized when the highly irregular comet body was finally resolved 10.5 years after launch.

The shape of comet 67P is composed of two unequal lobes with a deep neck area connecting the two. This added complexities in the gravity field, the non-uniformity of material and light coming off the surface, and the geolocation mapping. The combination of these factors along with the anticipated difficulties of orbiting a low gravity, dynamic body that creates its own ever-changing environment resulted in truly unique mission challenges that required equally unique operations solutions.

As a member of the Jupiter family comets (JFCs), 67P's highly elliptical orbit takes it just past Jupiter's orbit (5.684 au) and almost as close to the Sun as Earth (1.246 au) every 6.45 years. The result is a periodic transformation between a dormant, dark, frozen nucleus near aphelion and a spectacularly dynamic display of sublimating gas and ejected dust near perihelion. This strong variation in activity drove many Rosetta mission attributes and constraints, adding to the already complex system that included the Rosetta orbiter with 11 instrument packages and the Philae lander with its own set of 10 instruments.

As the comet activity evolved as it circled the Sun, the Rosetta and Alice operations processes evolved through the mission to provide a system that could meet the ever-changing needs. Tools were redesigned, interfaces changed, and processes updated to optimize science return.

After launch, the circuitous path of Rosetta included several orbits in the inner solar system including several gravitational assists from Earth and Mars, the last of which sent Rosetta to the outer solar system on a rendezvous path with 67P. Along the way instruments were able to make observations of Earth, comet LINEAR, comet Tempel 1 during the Deep Impact event, Mars, asteroid Steins, and asteroid Lutetia, which allowed testing out flight systems and ground processes. Shortly after waking from an unprecedented pre-prime mission 2.5 year hibernation period that lasted through aphelion without any communication with ground stations on Earth, Rosetta began studying comet 67P in a manner never before attempted at a comet. Where previous missions had studied comets from flybys of several 100's km away traveling at relative velocities of 10's of km/s or performing a brief surface impact, Rosetta's two year encounter period was often spent at 10's of km from the comet nucleus with relative velocities of meters/second. A slow, controlled surface impact concluded the historic mission adding to the high surface resolution measurements that the Philae lander acquired at the beginning of encounter. Table 1 details some of the important Rosetta mission and Alice instrument events.

*Table 1*: Major Mission and Alice Events

| Date | Event |
|---:|---|
| 2004 Mar 2 | Launch |
| 2004 Apr-May | Observations of comet LINEAR |
| 2005 Mar 4 | Earth flyby #1 |
| 2005 Jul 4 | Observations of Deep Impact |
| 2007 Feb 25 | Mars flyby |
| 2007 Nov 13 | Earth flyby #2 |
| 2008 Jul 19 | ➢ **Alice upset/reset event #1** |
| 2008 Sep 5 | Asteroid Steins flyby |
| 2009 Sep 29 | ➢ **Alice upset/reset event #2** |
| 2009 Nov 13 | Earth flyby #3 |
| 2010 Jul 10 | Asteroid Lutetia flyby |
| 2011 Jun 8 | • Spacecraft hibernation entry (4.5 au) |
| 2012 Oct 3 | Aphelion (5.3 au) |
| 2014 Jan 20 | • Spacecraft hibernation exit (4.5 au) |
| 2014 Mar 18 | ➢ **Alice 1st power on after hibernation** |
| 2014 Jun 6 | ➢ **Alice begins continuous powered operations** |
| 2014 Aug 6 | Orbit insertion at comet 67P, start of encounter period |
| 2014 Nov 12 | • Lander delivery (3.0 au) |
| 2014 Nov 15 | • Lander hibernation begins |
| 2015 Mar 28 | • Spacecraft star tracker safe mode – instruments off |
| 2015 Apr 15 | ➢ **Alice returns to nominal operations after safe mode** |
| 2015 May 10 | Equinox (N vernal, S spring) |
| 2015 Jun 13 – Jul 9 | • Intermittent lander contacts |
| 2015 Aug 13 | Perihelion (1.2 au) |
| 2015 Sep 4 | Solstice (N winter, S summer) |
| 2016 Mar 20 | Equinox (N spring, S vernal) |
| 2016 May 28 | • Spacecraft star tracker safe mode – instruments off |
| 2015 Jun 1 | ➢ **Alice returns to nominal operations after safe mode** |
| 2016 Sep 30 | End of mission (3.8 au) |
| ➢ **Alice Event**; • Spacecraft or Lander event | |

In Section 2 we describe the Alice Ultraviolet Spectrograph design and science objectives. In Section 3 we describe the operations of Alice and the various instrumental and mission constraints that affect the operations. Section 4 covers the different types of Alice observations, supporting observations by other instruments, and the special issues of regarding making observations while orbiting an active comet. In Section 5 we describe the broader context and design of Rosetta mission operations. Section 6 provides some lessons learned and additional information.

## 2 The Alice Ultraviolet Spectrograph

### 2.1 Alice Design

Described below is the design of the *NASA*-funded Rosetta Alice instrument aboard the ESA Rosetta asteroid flyby/comet rendezvous mission. Most of the text in the Alice Design section is from Stern et al., 2007 [1], but repeated here for stand-alone completeness of this document.

Alice is a lightweight, low-power, and low-cost imaging spectrograph optimized for cometary far-ultraviolet (FUV) spectroscopy [1]. It was the first of a family of similar instruments currently including Alice on New Horizons, LAMP UVS (Lyman Alpha Mapping Project Ultra Violet Spectrograph) on Lunar Reconnaissance Orbiter (LRO), UVS on Juno, and the expected UVS instruments on the upcoming JUICE and Europa Clipper Missions. Rosetta Alice was the first UV spectrograph to study a comet at close range. It was designed to obtain spatially-resolved spectra in the 700–2050 Å spectral band with a spectral resolution between 8 Å and 12 Å for extended sources that fill the slit.

An opto-mechanical layout of Alice is shown in Figure 1. Light enters the telescope section through a 40 × 40 mm$^2$ entrance aperture at the bottom right of Figure 1 and is collected and focused by an f/3 off-axis paraboloidal (OAP) mirror onto the entrance slit and then onto a toroidal holographic grating, where it is dispersed onto an imaging photon-counting microchannel plate (MCP) detector that uses a double-delay line (DDL) readout scheme [2]. The two-dimensional (1024 × 32)-pixel format, MCP detector uses dual, side-by-side, solar-blind photocathodes: potassium bromide (KBr; for λ < 1200 Å) and cesium iodide (CsI; for λ > 1230 Å) [3]. The measured spectral resolving power (λ/ Δλ) of *ALICE* is in the range of 70–170 for an extended source that fills the instantaneous field-of-view (IFOV) defined by the size of the entrance slit.

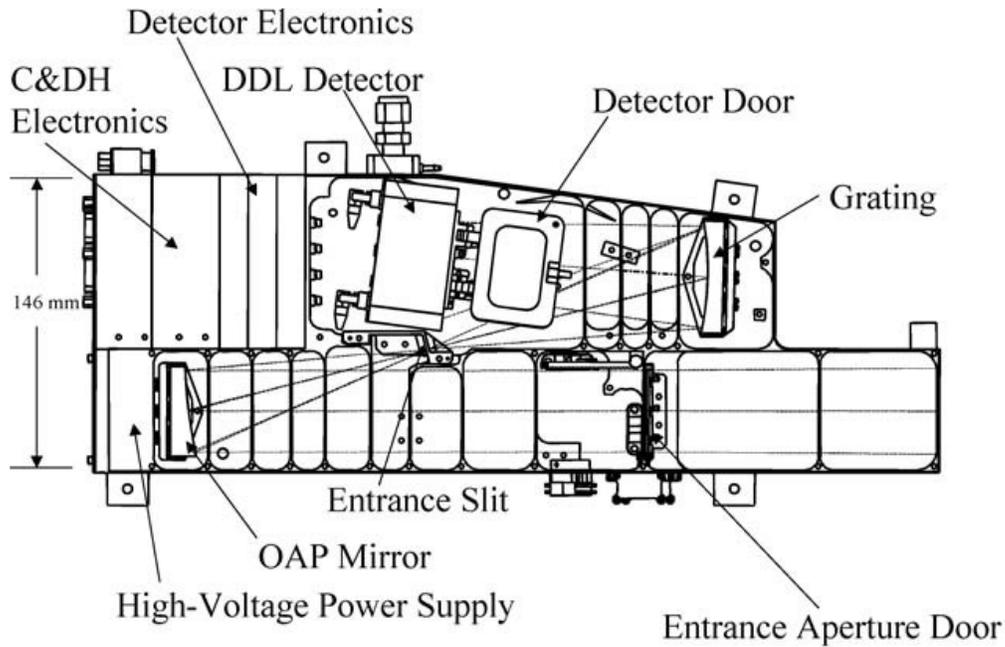

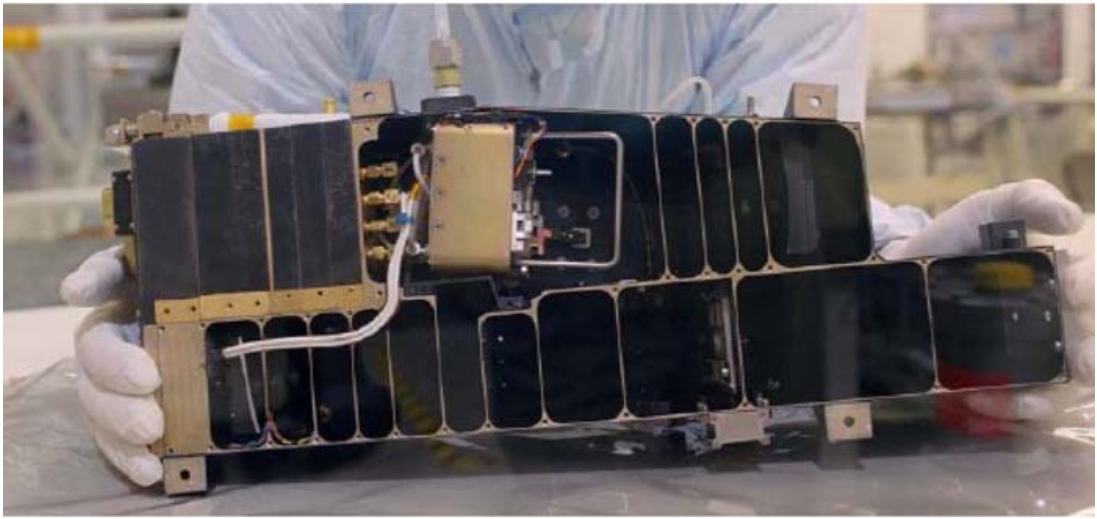

*Figure 1.* (a) The opto-mechanical layout of Alice. (b) A photograph of the Alice flight unit. [1]

The slit design is described as a "dogbone"; it is 5.53° long in the spatial dimension, the central 2° of which (called the "narrow center") is 0.05° wide in the spectral dimension, and the 2° "wide bottom" and 1.5° "wide top" portions of the slit are 0.1° wide in the spectral dimension, shown in Figure 2. The pinhole feature at the top of the slit was never detected in flight.

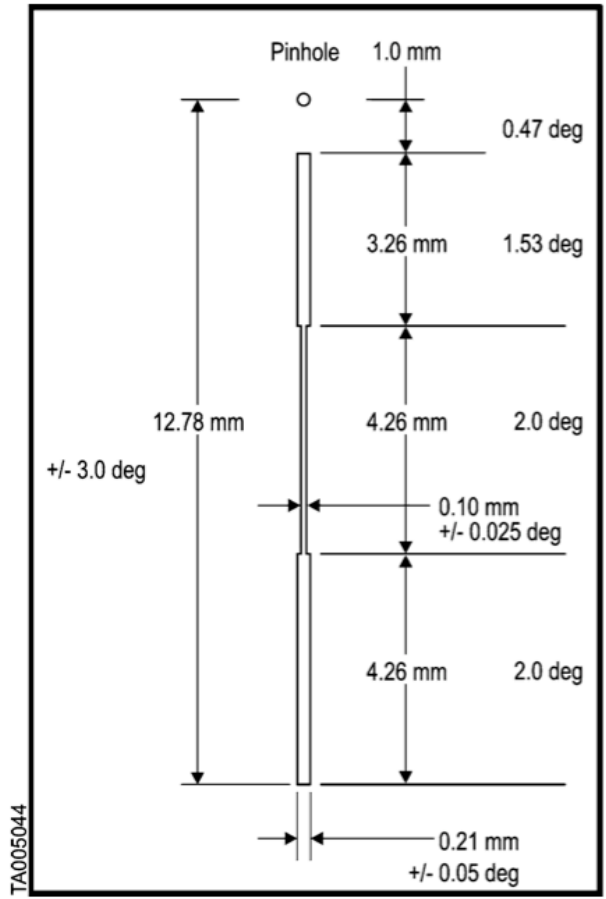
*Figure 2.* The Alice entrance slit design [1]

To capture the entire 700–2050 Å bandpass and 5.53° spatial field of view (FOV), the size of the detector's active area is 35 mm (in the dispersion direction) ×20 mm (in the spatial dimension), with a pixel format of (1024 × 32)-pixels. The 5.53° slit-height is imaged onto the central 20 of the detector's 32 spatial channels; the remaining spatial channels are used for dark count monitoring. The pixel format allows a Nyquist sampled spectral resolution of~3.4 Å and a spatial resolution of 0.3°.

The detector electronics amplify and convert the detected output pulses from the MCP Z-Stack to pixel address locations. Instead of discrete physical pixels as in a CCD detector, the time of arrival of a charge at both ends of the orthogonal delay lines is used to determine the spectral "pixel" on which a photon landed. The sensitivity of the detector electronics to convert a photon event into an output pulse can be adjusted by changing the detector high voltage level. Only those analog pulses output from the MCP that have amplitudes above a set threshold level, also referred to as the discriminator level, are processed and converted to pixel address locations. For each detected and processed event, a 10-bit *x* address and a 5-bit *y* address are generated by the detector electronics and sent to the Alice command-and-data handling (C&DH) electronics for data storage and manipulation. In addition to the pixel address words, the detector electronics also digitizes the analog amplitude of each detected event output by the preamplifiers and sends this data to the C&DH electronics. Histogramming this "pulse-height" data creates a pulse-height distribution function that is used to monitor the health and status of the detector during operation. A built-in "stim-pulser" is also included in the electronics that simulates photon events in two pixel locations on the array. This pulser can be turned on and off by command and allows testing of the entire Alice detector and C&DH electronic signal path without having to power on the detector high-voltage power supply. In addition, the position of the stim pixels provides a wavelength fiducial to account for shifts that can occur with operational temperature changes.

## 2.2 Alice Science Modes

The Alice instrument has three modes of taking data: histogram, pixel list, and count rate. The first two acquisition modes use the same event data received from the detector electronics, but the data are processed in a different way. The third acquisition mode only uses the number of events received in a given period of time – no spectral or spatial information is provided.

In *histogram mode*, the instrument continuously collects detected events by pixel location throughout the exposure, much like how a CCD counts and accumulates detected photons. During the course of a histogram observation, each detected event is saved into memory, where the memory locations map one-to-one to pixels on the detector. In this way, an integrated spatial-spectral image is accumulated over the commanded exposure time. This mode is useful for observations with high count rates, and was the most commonly used mode during the mission. The resulting images are the same format as the detector, 1024x32 pixels, with the long axis corresponding to wavelength and the short axis is the spatial dimension. Since the outer edges of the image are an "unexposed" part of the detector, the software places the pulse height array in this region, and it also provides monitoring of background/cosmic ray and dark counts. The histogram values (counts) saturate at 65535 counts.

In *pixel list mode*, the instrument serially records the x-y location of each photon and inserts time tags ("hacks") in this list at regular, user-definable intervals to provide the timing information for the photons. Due to data volume considerations and the time required for memory readout, this mode was used primarily at low count rates. The maximum number of entries is 32767 (including photons and time tags), but may be less depending on the combination of brightness of the target and the exposure time of the particular observation. Once the memory is filled to this limit, no more data are recorded until the exposure is stopped and the memory is read out. This mode was employed primarily during the Bi-Monthly Cross Slit Scan Calibration due to the need for time resolution to detect when the star crossed the edges of the slit.

In *count rate mode* (also called "photometer" mode), the spatial and spectral information for each photon is ignored. The number of photons detected in each user-defined time interval are summed and saved as a single number in memory. This mode was not widely used, but could be useful for observing very bright ultraviolet stars. The maximum number of intervals is 32678, but may be less depending on the combination of time interval and total exposure time of the particular observation. (Note that the number of intervals is independent of the brightness of the target.) The summed counts per time interval saturate at 65535 counts. [4]

## 2.3 Alice Science Objectives

The scientific objectives of the Alice investigation were defined as follows and further details found in [1]:

1. Search for and determine the evolved rare gas content of the nucleus to provide information on the temperature of formation and thermal history of the comet since its formation.
2. Determine the production rates of the parent molecule species, $H_2O$, $CO$ and $CO_2$, and their spatial distributions near the nucleus, thereby allowing the nucleus/coma coupling to be directly observed and measured on many timescales.
3. Study the atomic budget of C, H, O, N, and S in the coma as a function of time.
4. Study the onset of nuclear activity in ways Rosetta otherwise cannot.
5. Spectral mapping of the entire nucleus of 67P at FUV wavelengths in order to both characterize the distribution of UV absorbers on the surface, and to map the FUV photometric properties of the nucleus.
6. Study the photometric and spectrophotometric properties of small grains in the coma as an aid to understanding their size distribution and how they vary in time.
7. Map the spatial and temporal variability of $O^+$, $N^+$, $S^+$ and $C^+$ emissions in the coma and ion tail in order to connect nuclear activity to changes in tail morphology and structure near perihelion.

# 3 Alice Operations and Constraints

## 3.1 Contamination

The material coming off the comet posed a serious mission risk as the fine dust and gas could coat instrument optics and solar panels which could lead to severe science degradation, decrease spacecraft power production, and prevent spacecraft attitude and location knowledge determination. Furthermore, during periods of high comet activity there were unpredictable strong outbursts that quickly propelled concentrated jets of material outward. Large pieces of material were seen orbiting the comet and at times images showed that Rosetta was flying through a blizzard of dust particles as shown in Figure 3.

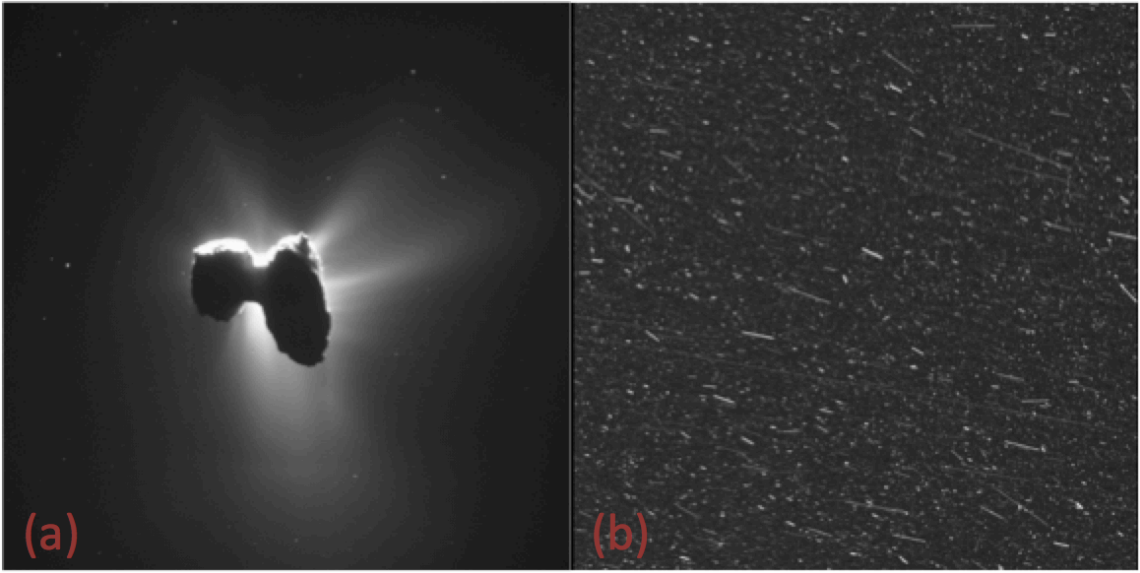

*Figure 3*. (a) Dust and gas escaping the comet nucleus on March 27 2016 [5]
(b) The dust environment around Rosetta captured by Rosetta's OSIRIS (Optical, Spectroscopic, and Infrared Remote Imaging System) camera on July 6 2015 [6]

It was difficult to strike the right balance between what was perceived to be a safe distance from the material coming off the comet and being close to the nucleus to obtain high resolution measurements and significant counts for the dust and gas instruments. For Alice, the local gas pressure and dust count measurements never reached sustained levels that would cause great concern of contamination, and the levels were always below limits that would have triggered an Alice safing event that closes the aperture door and turns off the detector high voltage. Regular calibration observations with the Alice instrument through the escort phase tracked sensitivity degradation, which was minimal, due to contamination and other sources.

However, one system affected by the dirty environment was the spacecraft star trackers that occasionally would get confused by the bright dust particles and lose attitude knowledge. That resulted in two spacecraft safe events that halted all activities to regain attitude knowledge, and there were several other close calls (as further detailed in Table 1). Since such spacecraft safety events could jeopardize the entire mission, the program erred on the side of caution and usually kept a healthy margin on the distance to the comet based on activity. This typically helped Alice science because the spectrograph's long slit (5.53°) at these larger distances allowed Alice to have some spectral elements on the nucleus and others looking off-limb at the escaping gases in tandem (as a "ride-along") with observations designed by other instrument teams that preferred surface pointing (see the Figure 4 example). Another factor that possibly helped preserve the Alice throughput is that the instrument operated in continuous decontamination mode for most of the encounter after verifying it did not impact instrument calibration. This meant that the mirror and grating heaters were constantly enabled to keep those optical elements warm so they would not be cold traps by driving off any deposited contamination. However, given that there were very few issues for components on the Rosetta orbiter through encounter phase demonstrates that encountering high concentrations of material and large destructive chunks of material was rare even though the environment was much dirtier than typical when orbiting less active bodies such as planets, moons, and asteroids.

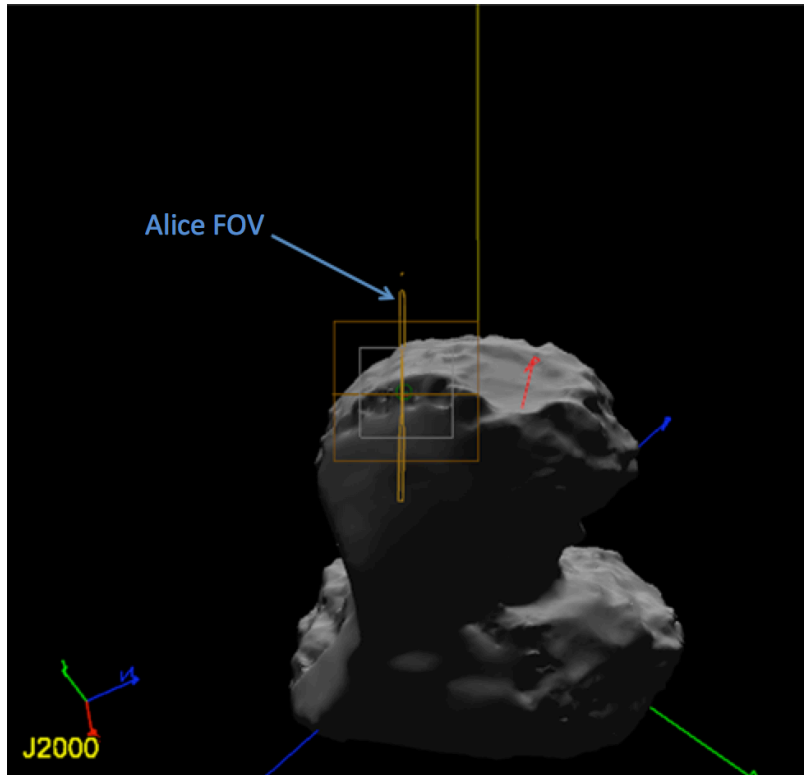

*Figure 4*. Example of an OSIRIS targeted surface observation (lander search) that could be used by Alice as a ride-along to view off-comet gasses [7].

## 3.2   Gain Sag

Adding some unique constraints to the Alice operations, the instrument has an un-scrubbed microchannel plate (MCP) detector. The scrubbing process exposes the detector to UV light reducing the pixel charge, a conditioning technique that "burns in" the pixels, eventually stabilizing the detector response. However, a scrubbed detector imposes significant constraints in pre-flight ground handling including requiring being kept in constant vacuum, which was not an option in this case as the Alice bandpass in the EUV required an open-faced detector. The consequence of flying an un-scrubbed detector is that its response continually decreases through the mission from exposure to UV light, a typical characteristic of MCPs called "gain sag". The gain sag is also spatially dependent as a function of the total fluence on a given area of the detector, leading to non-uniform sensitivity degradation across the detector. For example, the varying brightness of the comet surface, the observation of bright UV stars on specific detector rows, or the constant exposure to Lyman-alpha emission from the interplanetary medium will cause some parts of the detector to degrade differently from other parts (see Figure 5). The degradation was tracked, accounted for in the data calibration, and periodically compensated for by changing detector settings such as high voltage level (a detector sensitivity adjustment) and discriminator level (a signal cutoff adjustment). Characterization of the detector performance was conducted periodically using moderately bright UV stars as calibration sources.

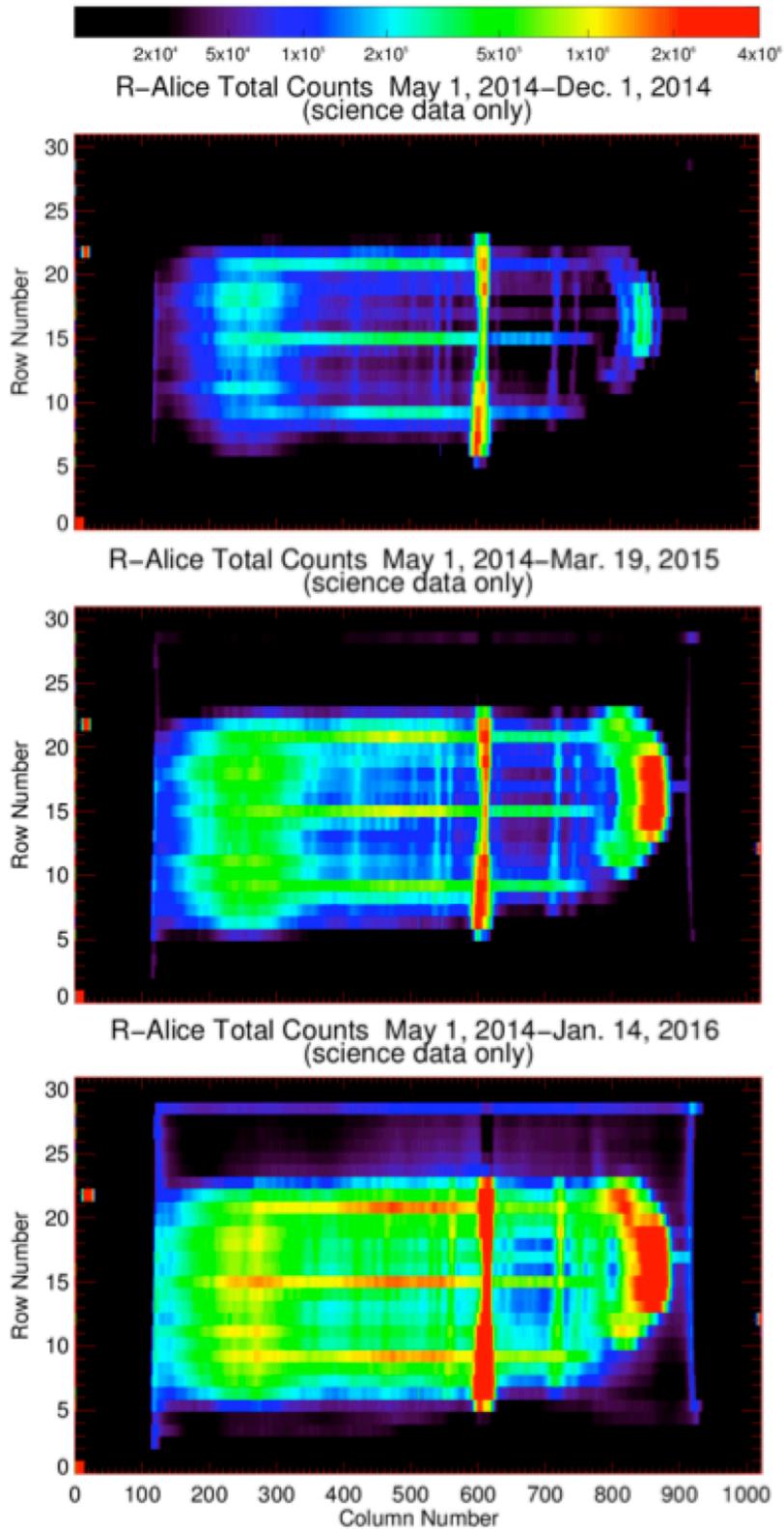

*Figure 5*. The progression of detector science counts through encounter. The concentration of counts near column 600 (1215 Å) is the Ly-α line and the region around column ~850 (~1025 Å) corresponds to the chameleon artifact region. The high concentration of counts in rows 9, 15, and 21 are due to stellar calibrations. Wavelength decreases to the right, i.e., to higher row number.

The anticipated long term gain sag drove strategic planning for the detector usage to ensure that the detector would maintain adequate sensitivity throughout the mission in critical bands and detector regions to satisfy the science goals. This initially drove analysis-intense observation constraints of conducting only high priority comet measurements and actively avoiding exposure to relatively bright UV stars.

### 3.3 Bad Dogs and Chameleon Contribution to Gain Sag

Some stars were so bright in the UV that they could cause excessive count rates on the Alice detector. These UV-bright stars were nicknamed "bad dogs". The Alice team would receive the predicted spacecraft attitude and location information and use it do determine when bad dog stars would cross the Alice FOV plus a margin to account for pointing uncertainty. The team would then adjust instrument commanding to suspend the Alice observations and close the aperture door around the forecasted bad dog encounters. Making the process even more complex was that the planned spacecraft attitude could change late in the product development process due to other instrument teams adjusting their prime observations or by the mission flight dynamics team adjusting the comet orbit plans. This would require agile monitoring and response from the Alice team to quickly (sometimes as short as within a couple days) identify and make corresponding Alice observation changes.

As the mission progressed, it was determined that the detector gain sag rate was lower than initially projected, which allowed the gain sag avoidance techniques to become progressively less conservative. By the end of the mission Alice was observing almost continuously and bright UV stars were no longer manually avoided. Instead of avoidance planning, a detector count rate monitor with a configurable threshold was relied upon to react when any too-bright stars came in the FOV, autonomously safing the instrument by temporarily pausing the observing and closing the aperture door. The momentary exposure to the bright source did not significantly contribute to the overall detector gain sag and was a welcome tradeoff between some loss of exposure time in exchange for significantly reducing the labor-intensive star avoidance planning.

The other primary contributor to bright safety events came from a sporadic issue referred to as the "chameleon", which is thought to be due to charged particles entering the instrument, getting through the detector electron grid, and causing strange, high-count measurement artifacts (see Figure 5) [8]. Because the chameleon had strong temporal variations, usually the bright safety action of temporarily closing the aperture door and waiting the safety timeout period (typically 10 minutes) was sufficient to deal with the occasional high-count rate chameleons.

### 3.4 Alice Observing Styles

Throughout the mission Alice would observe during both Alice-planned (termed "prime") observations as well as observations planned by other instrument teams (termed "riding along"). Even though many observations designed by other teams weren't optimized for Alice science goals, some science return was always possible when observing the comet and its surroundings. However, potential science return always had to be balanced with available spacecraft resources, development effort, and the expected gain sag impact. Alice observing consisted of two main styles. The first was continuous observing where histograms were taken one after another (with short ~40s overhead for image readout between exposures) for hours at a time, not correlated to a specific scene or pointing changes. This was simple to implement, but could lead to difficult to use data such as when histograms bridged changes in spacecraft staring and scanning motions or major scene changes. The other type was to manually time histograms to correlate their start/stop times with specific viewing conditions. Although this technique could often return better science, this style had a high cost in terms of development effort and commanding complexity. This was especially true when timing histograms during ride alongs because other instrument teams would often change their pointing plans during the development period, which usually required corresponding adjustments to Alice observations. At the beginning of the mission the Alice team focused on its prime science refraining from observing during pointing planned by most of the other teams. However, as the mission progressed, ride alongs increased when data volume and other resources were available to a point where Alice was observing almost continuously. This was especially true after it was evident that the gain sag progression rate was less than expected and software tools were created to automate observation development based on planned pointing characteristics.

## 3.5  Aperture Door Usage and Power Cycling

Alice instrument operations are unlike a lot of other space instruments in terms of power cycling and mechanical door operations. Due to the risk of power up failures, the action of power cycling space electronics has long been avoided and on some missions only done when forced to because of a failure on one string of a redundant system. By necessity this operational method was not an option on Rosetta. Power constraints during certain periods of the mission required instruments and some non-essential spacecraft components to alternate powered activities leading to many unavoidable component power cycles. In fact, during an unprecedented 2.5 year hibernation period when Rosetta was at the outermost part of its orbit, which took it out to the orbit of Jupiter, it only had enough solar power to have a few vital heaters and electronic units on [9]. There was no contact with the spacecraft, not even beacons during that hibernation period. An onboard timer was set to tell Rosetta to wake up on January 20, 2014 when there would be enough power to turn some systems on and contact Earth. Everyone was on the edges of their seats on hibernation exit day to say the least and amazingly all systems turned on and proceeded with the comet encounter 10.5 years into the mission.

During the escort phase (August 2014 – September 2016) standard Alice operations were to power cycle the instrument on a weekly basis. An Alice upset/reset event occurred twice during the cruise period getting to the comet (see Table 1 for the dates). The anomaly would power cycle Alice and restart in a default state defined by hard-coded parameters, and although it would return to the nominal activity timeline, science data quality collected afterward could be negatively impacted until manual adjustments were commanded to several operational parameters. The root cause of the upset/reset events was not identified, but it was thought that

Rosetta Alice Ultraviolet Spectrograph Flight Operations and Lessons Learned        13

periodically power-cycling the instrument perhaps could prevent such a problem from occurring or would return Alice to its nominal state if an upset occurred. With some power cycles already required and four separate EEPROM flight software images to boot from, it was decided that the risk would be lower to perform regular power cycles rather than being on continuously. Additionally, to mitigate the impact of an upset/reset event, activities were split into small independent segments and several operational parameters were reasserted twice daily.

The Alice aperture door operations were similarly different from most other space mechanical systems. Whereas most mechanical systems, and doors especially, are designed for one time use, the Alice aperture door was designed to be opened and closed many times, rated for 10,000 such "flaps", which gave a factor of 2 margin relative to ground tests. Alice relied on a robustly designed aperture door mechanism proven by extensive ground qualification testing and subsequently by later versions of the instrument. The LAMP UVS, which is a version of Alice on the Lunar Reconnaissance Orbiter mission, has performed over 95,000 cycles. As a safeguard, Alice also had a failsafe door that could be opened permanently to retain some science ability if the aperture door failed shut. Driving all the door cycles was the instrument's susceptibility to contamination and detector gain sag. To avoid contamination of instrument optics from spacecraft thruster byproducts, the aperture door was closed during, and for 30 minutes after, every spacecraft propulsive maneuver. To avoid contamination from cometary material the aperture door was also closed for gaps in observations of more than 15 minutes through most of the encounter and even shorter gaps when the comet was especially active. Monthly door performance tests confirmed the mechanism exhibited no measurement trends that would cause concern for continued regular use.

## 4 Alice Observations

### 4.1 Orbiting a Comet

There was a preliminary encounter phase timeline outlining a strawman plan of Rosetta's orbits around the comet, but due to the unique cometary environment it took some time for the mission teams to get used to operating around a comet and refine the plan. In addition to the standard engineering and environment models needed for interplanetary travel, many unusual factors needed to be accounted for including: orbiting a low gravity body with an asymmetric gravity field created by the two lobed comet shape, variable aerodynamic drag due to temporal and spatial variability of cometary activity, navigation with star trackers in an environment with high visual dust confusion, and contamination risks. The orbits varied quite a lot to account for these constraints and to accommodate the requested observational conditions, but the following describes the typical obit types:

The spacecraft flew low (<32 km radius) circular orbits when the comet was less active. This allowed the instruments to obtain high-resolution surface information when the contamination risk was low, and to increase source flux for the in situ instruments.

As comet activity increased, the spacecraft orbit was expanded to several hundred kilometers to a maximum radius of about ~600 km around perihelion as a safety precaution, but to accommodate science needs, the mission tried to keep the orbit height as low as safety would allow. Bound orbits were not used when orbiting above ~32 km both due to the difficulty of balancing acceleration factors when orbiting a low-gravity body during periods of higher comet activity. Above that height, the spacecraft transitioned to joined segments of hyperbolic arcs that typically resulted in full orbits that were roughly circular shape. This maintained regular observation conditions for science planning while also providing the security of a safe trajectory that would avoid collision with the comet if control of the spacecraft were lost [9]. The orbits typically were aligned with the terminator to minimize gas drag on the spacecraft and avoid eclipses. The orbit orientation allowed the solar panels to be continuously pointed to the Sun and resulted in the minimum surface area pointing in the comet direction. However, angle offsets were occasionally used to obtain a variety of illumination conditions.

Close comet flybys, as close as 7 km from the surface, were used for high surface resolution measurements.

Distant "excursions" to 1000 km in the tail direction and 1500 km in the Sun direction were used to measure the local magnetic field properties.

Finally, the low gravity comet body allowed for some unusual orbit types that would have required an exorbitant amount of fuel to accomplish if Rosetta orbited a larger body requiring higher relative speeds. For instance, sharp turns were used for special activities including the Philae Lander release trajectory and Rosetta swept back and forth over a single area above the surface in an effort to locate the Philae Lander (see Figure 6). Because of the low orbital speeds of meters per second, a large change in direction actually could involve in a relatively low change in momentum, and thus, not a significant use of fuel.

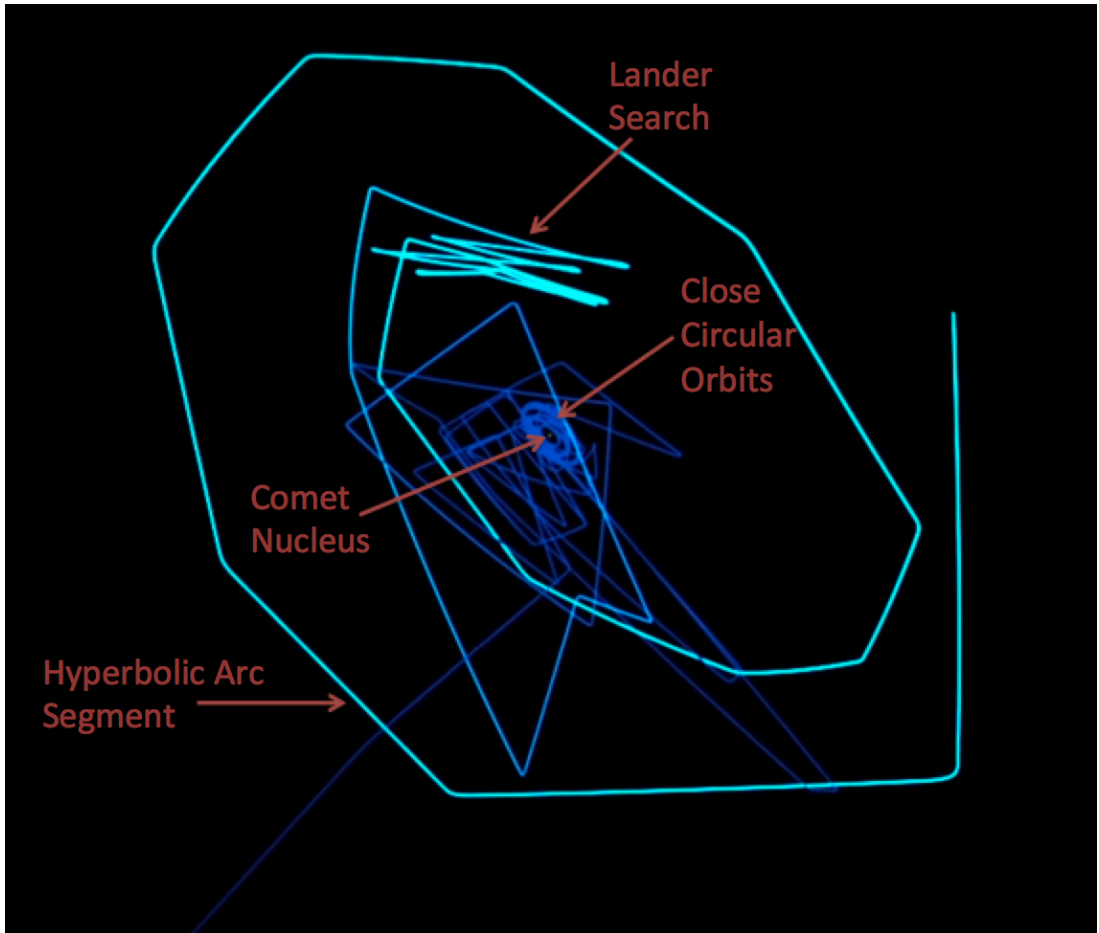

*Figure 6*. Rosetta orbit diversity example [10]. The dark blue color displays the approach beginning July 31 2014 and early encounter close orbits. The light blue color displays more distant orbits around perihelion through August 9 2016 and the lander search.

## 4.2 Pointing Uncertainty

Another unique challenge this mission had to contend with while planning observations was fairly large pointing uncertainties. As is the case for most space missions, the spacecraft flight dynamics team provided the science teams many orbital parameters to use for planning observations. However, unique to this mission was the task of understanding the craft's relative location and attitude with respect to the comet in light of significant non-gravitational forces. The spacecraft inertial location and attitude was known very well from the star tracker data and enabled communication with ground stations on Earth and precise pointing at stars for instrument calibrations. However, there were often large uncertainties for the spacecraft pointing relative to a comet reference point such as nadir, limb, or surface locations and this had a significant impact on observation planning at several levels. The pointing uncertainty source came from the combination of environment and comet model uncertainties, thruster efficiency uncertainties for upcoming orbit correction maneuvers, and the quality of the comet-relative location knowledge as ascertained from navigation images of the comet.

The pointing uncertainty continually changed, being best just after the current spacecraft location was determined and updated on the spacecraft computer and then it would grow until the next update as the uncertainties compounded in the models. The projected comet-relative pointing uncertainty range was typically between 0.1° – 4.0° for most of the mission, but when the spacecraft traveled very close to the comet surface (under 10 km) the uncertainty forecast could grow much larger up to ~16°. In practice it was found that the difference between the desired and actual pointing could occasionally grow large for short periods of time, however conservative factors in the models typically resulted in a large overestimate for the pointing uncertainty, typically by a factor of 0.5 or more.

Even if observations were planned using pointing uncertainty values of half the forecasted level, the scale of the uncertainties was very difficult for some of the instrument teams to incorporate in their planning. Negotiations for prime observation slots by the remote-observing instrument teams often prioritized slots with low pointing uncertainty values to reduce observation smear. However, Alice planning was able to be more flexible than that for other instrument teams regarding pointing constraints and acquired a lot of prime observations with high pointing uncertainty periods. Making use of these slots was made possible since the Alice FOV was much larger than most of the other instruments (5.53° long) and it took relatively long exposures (5-10 minutes) making observations of specific locations less important than getting good regional coverage. Also, several Alice observation types were off-comet, including inertial or 360° Great Circles that were not impacted by the comet-relative pointing uncertainty. This made Alice observing much less dependent on the pointing uncertainty values compared to the others and Alice was able to put many of the worse pointing uncertainty blocks to good use.

### 4.3 Observation Types

Specific pointing could be developed once the orbit type was defined and the trajectory and sub-spacecraft parameter details frozen, such as orbit height, phase angle, latitude, longitude, and pointing uncertainty. After negotiations for observation time were completed, the instrument teams turned their attention to implementing the plans. Each instrument team would tailor the pointing during their prime observations to optimize science return. Teams could verify and iterate the pointing products using mission simulator tools, then after they were delivered they were verified and approved by the Rosetta Science Ground Segment (RSGS) group at ESA's European Space Astronomy Centre (ESAC) in Spain and the Rosetta Mission Operations Center (RMOC) group at the European Space Operations Centre (ESOC) in Germany. There were quite a range of observation types implemented by the various instrument teams, but most were variations on these:
- Staring at the surface with the comet rotating underneath the spacecraft
- Matching the comet's rotation rate to track a surface location
- Scanning across the surface
- Staring at one of the comet limbs
- Pointing off comet

Below are details of some of the variants of these observation types that Alice utilized. For reference the observation names are also used in Alice data nomenclature, which is discussed further at the end of the paper in the Alice Logbook Section.

## 4.4 Alice Observation Types: Gas Observations

A primary Alice science objective was to characterize the gasses coming off the comet to better understand the solar system's primordial constituents and environment. This required measuring the different gas types, their relative abundances, and how they varied through the evolution of the comet during its orbit around the Sun. To do this, a variety of observation types were designed to view the gasses against a dark background (either looking off comet or a shadowed region on the comet) where the comet surface reflectance wouldn't interfere with the measurements. The styles varied and were refined based on distance to the comet, gas abundance, and the understanding of the comet environment.

*Volatile Abundance Campaign.* A primary style of gas observation used for Alice and other instruments was named Volatile Abundance Campaign (VAC), which was designed to measure illuminated gasses against the deep space background (a "Deep VAC" was a long exposure version of such pointing). This was a stable staring observation where the Alice slit would be positioned such that it was approximately half on the comet surface for location context and half off the sunward limb, though the exact position varied due to pointing error and negotiations with other instrument teams.

*Inner Coma Raster.* When the comet was more active, Inner Coma Raster observations would be used, which would stare in several locations stepping off the comet in a radial line, typically sunward, to track how the gas abundance falls off with respect to distance from the comet.

*Great Circle.* Since Rosetta was near the center of the comet's coma throughout the encounter period, the Alice instrument took full 360° "Great Circle" observations to measure the distribution of the gasses it was immersed in and how they changed throughout the orbit around the Sun.

*Night Stares.* These observations were designed to look at the illuminated gas between the spacecraft and the comet surface either on the night side of the comet or over large shadowed regions, which had the useful effect of cutting out the sky background.

*Coma Ride-Alongs.* The Alice team also rode along on similar observations planned by other instrument teams such as the OSIRIS-led VACs (termed Diurnal VACs) that scanned back and forth along the sunlit comet limb looking for concentrations of activity and the VIRTIS (Visible and Infrared Imaging Spectrometer) instrument Snowflakes that were a combination many off-comet stares.

*Stellar Occultations.* Although observing stellar occultations was always part of the observation concept for Alice and other instruments to measure coma gases in absorption, planning those proved to be unfeasible due to the uncertainties in the predicted spacecraft orbit and attitude. However, as UV stars were monitored passing through the Alice FOV, it was realized that it was still valuable to observe stellar appulses even if the star didn't get to the absolute lowest layers of the coma just above the nucleus. The Alice team was able to model (within uncertainties) the tangential distances from the nucleus of various UV-bright O- and B-type stars to target those that could be observed through the column of gas near the sunward limb of the comet. The large pointing uncertainties and frequent pointing plan changes made planning these observations difficult, but when such observations were successful, the same star was observed again at a later time far away from the comet as a baseline measurement. These two spectra were compared to measure the level of attenuation in specific bands to characterize gas type abundance. The technique was initiated late in the mission after perihelion, so a time history of these measurements throughout the comet's activity cycle could not obtained, but $H_2O$ and $O_2$ abundances were measured from several of the 29 stellar appulse measurement sets [11].

### 4.5 Alice Observation Types: Surface Studies

A secondary Alice science objective was to characterize the comet surface reflectance in the UV and search for surface ices to better understand the comet's composition and morphological processes. For these investigations two types of observations of the illuminated comet surface were planned.

*Surface Center.* This style of observation would stare at the surface either directly nadir or with a defined offset from nadir and take continuous histograms as the comet rotated underneath the spacecraft with a period of 12 hours 24 minutes. If the comet's rotation axis was not aligned so that the scene would vary through the observation period, the observation type would be changed to a Surface Scan.

*Surface Scan.* This style would scan very slowly at 0.01°/s back and forth from the sunward limb to the anti-sunward limb (spacecraft Y-axis, perpendicular to the ALICE slit) or as far as could be scanned within an observation window when close to the comet.

*Surface Ride-Alongs.* Alice rode along on an assortment of other instrument observations especially on long stares and the Targets of Opportunity (ToO), where specific interesting surface features were stared at and most instruments recorded data to provide a comprehensive study of the area. There were a fair amount of observations optimized for other instruments that Alice did not typically ride along because they wouldn't yield high quality Alice data. For instance, observations that included high scan rates or short stares of less than 5 minutes for fast mapping of the surface would smear too much surface area to be of much use for Alice analysis. Alice typically used 5-10 minute histograms for surface observations and 10-20 minute histograms for gas observations to balance measurement signal and spatial resolution.

### 4.6 Alice Observation Types: Calibrations

It was very important for several Rosetta instruments including Alice to characterize their performance with regular calibrations. In addition to the typical instrument optics degradation and alignment changes, each Alice observation contributed to a non-uniform detector gain sag effect that needed to be tracked and accounted for. Moderately bright UV stars were used as stable calibration sources for Alice, however they typically were far off-comet and it would take a lot of time to slew to a calibration star's position. Since all the instrument teams preferred to spend as much time as possible observing the comet, calibration observations were coordinated to minimize off-comet pointing time. OSIRIS, VIRTIS, and Alice were able to combine stellar calibration requirements reducing the total number of instances.

The Alice instrument performed several types of stellar calibrations to track changes in performance:

*Bi-Weekly Stellar Calibrations* were used to measure the effective area (instrument sensitivity) by periodically staring at calibration stars and measuring changes in response for three representative locations on the detector.

*Bi-Monthly Flat Field Calibrations* were conducted as a raster scan along the length of the slit of a calibration star to measure relative sensitivity differences across the 2D detector. This set of bi-weekly stellar and bi-monthly flat field measurements allowed for the effective area changes to be applied across the full detector providing a method of correcting for detector gain sag on a per pixel basis.

*Bi-Monthly Cross Slit Scan Calibrations* were performed that consisted of raster scans perpendicular to the slit of a calibration star using pixel list measurements to precisely define the instrument field of view. This was important to be able to track and account for the instrument flexure due to changes in thermal environment throughout the mission.

*Bi-Monthly High Voltage/Discriminator Calibrations* were performed by staring at a calibration star and making observations with a set of detector high voltage (HV) levels and discriminator levels, settings that can be adjusted to change the detector sensitivity. These instrument settings could then be optimized to mitigate gain sag effects. The standard operational high voltage and discriminator value were changed several times during the course of the mission as shown in Table 2. These changes are taken into account in the calibrated data, but may be of interest if working with the raw Alice data.

*Table 2*: Nominal Alice High Voltage and Discriminator Settings for the Rosetta Mission

| Initial Settings | HV -3.8 kV, Discriminator 0.09 V |
|---|---|
| 2006 Dec 3 | HV -3.7 kV, Discriminator 0.34 V |
| 2007 Feb 23 | HV -3.8 kV, Discriminator 0.09 V |
| 2007 Sep 13 | HV -3.9 kV, Discriminator 0.09 V |
| 2014 Apr 2 | HV -4.0 kV, Discriminator 0.09 V |
| 2014 May 12 | HV -4.0 kV, Discriminator 0.45 V |
| 2015 June 25 | HV -4.1 kV, Discriminator 0.45 V |
| 2016 Feb 23 | HV -4.2 kV, Discriminator 0.45 V |

*Bi-Weekly Dark Calibrations* were also performed where histograms were taken to measure and account for the detector response with the aperture door closed.

## 4.7  NCIR and OCIR Context Images

Even to the trained eye, Alice histograms can be difficult to interpret, especially when viewing partially shadowed terrain and the variable concentrations and structures of gas and dust coming off the comet. To aid in the interpretation of histograms, it was very useful examine Navigation Camera (NavCam) and OSIRIS Wide-Angle Camera (WAC) images taken around the same time to provide context to the scene as shown in Figure 7. The timing of the images planned by the Spacecraft and OSIRIS teams were often not well correlated to important Alice observations. So in another example of cooperation within the mission to achieve the best science possible, the Alice team was permitted to plan additional NavCam and OSIRIS Images Requests (NCIRs and OCIRs) with the constraint that the image data volume was covered by Alice resource allocations. The NavCam was used solely for Alice context imaging until May 11th 2016, where data volume constraints started to significantly increase. At that point there was a transition to use OSIRIS WAC images, which could be compressed, saving a large amount of data volume to be repurposed as additional Alice science observations.

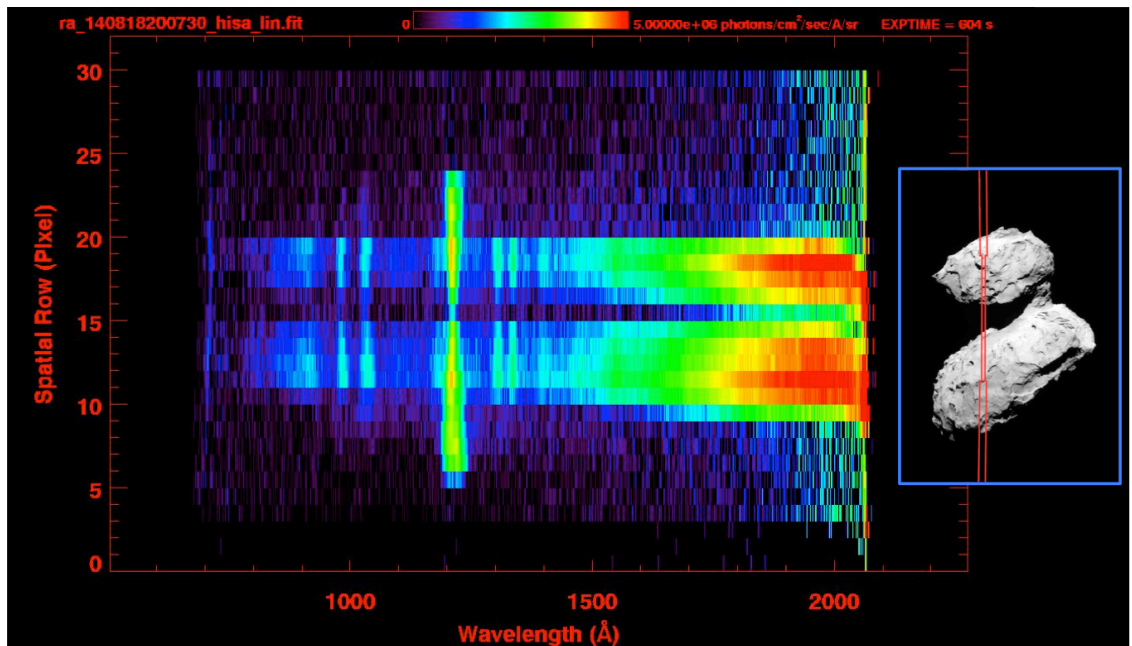

*Figure 7*. Example of an Alice histogram (left) of 67P and a NavCam context image [12] with an overlaid Alice FOV (right) acquired on 08/18/2014 (2014 DOY230). The context image can help identification of scene characteristics to correlate histogram features.

## 4.8  HEET Measurements

As a secondary science pursuit, Alice was periodically configured to measure the high-energy electron, or "HEET", component of the space environment. The Alice detector could sense penetrating electrons and would record them in the detector count rate data. This was accomplished at low data volume cost since the count rate is included in the Alice housekeeping data packet. HEET measurements were made with the Alice aperture door closed, the detector high voltage on, and at an elevated housekeeping rate of 1 Hz in an attempt to capture short-term electron density changes.

## 5 Mission Planning and Product Development

### 5.1 Mission Planning Coordination

At first glance the prospect of balancing the needs of the 21 Rosetta mission instruments seems daunting, especially with each based at separate institutions scattered across Europe and the US, in different time zones, and staffed by teams passionate about their anticipated science. Nonetheless, the mission was a resounding success and data were recorded to satisfy the Alice instrument science goals. This was achieved through tireless operations design, development, and planning support from all the instrument teams and the ESA science operations team. Additionally, the freedom to self-harmonize science planning within the Rosetta science community instead of being directed by management led to holistically balanced solutions.

An observation importance hierarchy between instruments was not always clear as mission science goals weren't requirements or ordered by priority. However, teams usually came to an agreement on their own when science priorities were made based on data set uniqueness and continuity. Often the teams made compromises to merge observation types for highly desired periods such as adding periodic stares during scans, adjusting scan rates, stare locations, etc. Fortunately, this task became more manageable as only a subset of instrument teams actively participated in resource negotiations including Alice, MIRO (Microwave Instrument for the Rosetta Orbiter), OSIRIS, ROSINA (Rosetta Orbiter Spectrometer for Ion and Neutral Analysis), RPC (Rosetta Plasma Consortium), VIRTIS, and sometimes the Philae Lander Team. The resources that the remaining instruments typically required were low enough that they were automatically granted or they were able to achieve their desired observations by riding along on the pointing designed by the other teams.

There were exceptions to the typical negotiation process, the biggest being how to balance the potential resource needs of the Philae Lander and its 10 instruments if and when it became functional again in the months after the landing. Sometimes, this led to resources and scheduling to search for the Lander or recover a Lander signal taking priority over other Rosetta spacecraft plans. That in turn resulted in late changes to the trajectory to optimize communication and view opportunities, added multiple-case planning, drove additional pointing constraints, and required resources to be reserved. Unfortunately, among the many contact attempts there were only brief additional contacts with Philae roughly 7 and then 8 months after landing (see Table 1 for the dates), the contacts were not sufficient to command extended additional lander science activities.

### 5.2 Long, Medium, and Short Term Planning Phases

Each instrument had its own master science plan that defined measurement goals, their relative importance, and observation methods to accomplish them. To achieve these goals each team would first make requests for the needed resources at the Long Term Planning (LTP) phase including spacecraft pointing, data volume, power, and telecommand count. This was done with a resource and request tracking spreadsheet typically covering 4 weeks of operations and was subdivided into short blocks of time defining spacecraft activities or blocks available for instrument teams to request. This typically resulted in 3-4 hour instrument observation blocks separated by 30-70 minute spacecraft activity blocks that instrument observing may or may not need to avoid due to propulsion and non-optimal pointing activities. Every observation block for the entire two-year encounter was reviewed and negotiated for by the instrument and spacecraft teams. Harmonization of plans took place with RSGS liaison scientists moderating the process from ESA's ESAC facility in Spain while the various instrument teams dialed in from around the globe. Harmonizing the observation requests often took a lot of time and compromises, and occasionally got heated when designs couldn't achieve everyone's needs. However, it was a testament to international collaboration and cooperation that most negotiations were attained smoothly and with enough time to implement. After harmonization the liaison scientists would hand off the plan to the Medium Term Planning (MTP) counterparts where commanding products would be built, refined, and validated. MTP periods were typically 4 weeks long, corresponding to the LTP planning spreadsheet periods. Finally there was opportunity for non-standard or emergency activities and changes to be added at the Short Term Planning (STP) phase, typically 1-week duration.

## 5.3  Evolution of Operations Planning

During the first 9 months of encounter the cometary activity was low and the spacecraft orbital trajectory planning was very stable, excluding the various Philae landing scenarios. This allowed for long development process durations, for example the LTP phase started 6 months before the execution of the planned activities, the MTP phase kicked off 2.5 months before execution, and there were several weeks for STP adjustments. This led to concurrent planning of several MTP periods, but plenty of time for leisurely planning. However, this luxury was short-lived as the comet became active and the orbital trajectory needed to be adjusted on a much shorter timeframe to keep a safe distance from the comet and particularly active areas such as the comet neck and the side of the comet pointed to the Sun where the gas and dust densities were higher. Planning phases were highly compressed after this point, but the mission was able to maintain the ability to plan observations that correlated to specific comet phase angles, locations, and other characteristics, which resulted in much better formulated datasets when compared to arbitrary pointing scheduling.

The remaining 16 months of the mission planning proceeded at a frantic pace with the LTP phase starting 2 months before execution and the MTP and STP phases merged and kicked off a month before execution. Unfortunately, last minute changes to pointing at the STP level by other instrument teams were not rare and usually required a scramble to accommodate in the commanding for Alice and other instruments. However, the Alice team was appreciative for the opportunity to make STP level changes to the pointing as well to optimize some observations, especially stellar appulses, which are discussed in the "Alice Observation Types: Gas Observations" Section.

The compression of the planning timeframe impacted all facets of operations from observation design, resource harmonization, implementation, development tools, product review timing, and product validation timing. It required that processes and systems be developed that could be efficient, adaptable, reliable, and modular. Among the Alice team, a wiki was used for efficient communication of product status and Subversion (svn) was used for product version tracking. Modular Python, Perl, and bash scripts were used for product development and verification that allowed for quick adjustments and testing as mission inputs and formats changed and new capability was needed. ESA mission simulation models (including ESA's Mapping And Planning Payload Science (MAPPS) software [7]), Satellite Tool Kit (STK) scenarios, and an Alice Engineering Qualification Unit were also used for product validation.

# 6 Mission Perspectives

## 6.1 Lessons Learned Summary

Cooperation and joint planning between teams lead to better overall science. ToOs, context images, star calibrations, and other cases of correlating several instrument measurements led to a more complete understanding of the comet processes and observed scenes.

Transparency in the operations planning for all instrument teams helped with understanding each other's goals and helped make the harmonization process more efficient. This was achieved by documenting observing styles, labeling observations in a standardized way, and collecting observation requests in one place for side-by-side review and harmonizing.

It is important to automate planning and commanding as much as possible to minimize manual effort. This leads to more efficient planning and reduces operational risk. It was found that rolling out capability incrementally and use of open source software quickened implementation of automation software. A recommendation for future Alice-like instruments would be to update flight software so it is flexible enough to create additional instrument sequences that combine sets of commands used regularly to reduce risk in command loads and make operations development more efficient.

In terms of personnel it was found that it was worth the resources to have someone who is working on operations development with intimate familiarity the systems and its needs with sufficient time to build and test operations tools to improve the process. There is also strong value in having a liaison between operations team and science team who can help translate requirements, constraints, jargon, etc. between the two teams.

Having a instrument flexible in pointing (made possible with a large FOV and long exposure times) meant the Alice instrument could take advantage of observation blocks that were of little use to other instruments, giving Alice more prime pointing than it would have otherwise secured.

Implementing an ongoing process of assessing operational constraints and assumptions help to identify areas to optimize. For instance, the gain sag rate was reassessed resulting in adjustments in Bad Dog avoidance and ride along observation planning.

The freedom to self-harmonize science planning within the Rosetta science community instead of using arbitrary scheduling or solutions being directed by management led to holistically balanced solutions and additional collaborations - not just compromises.

Being clear with communication and scheduling deadlines was very important to reduce issues for an international mission with team members spanning many time zones and native languages, but requiring many weekly meetings for planning coordination and quick-turnaround responses. One area of communications that could be improved on Rosetta was occasionally ESA mission software used for product development and modeling would change without ahead of time notice impacting ongoing development. This would require changes to interface and development software in a very short timeframe. Better transparency into mission software changes before they are implemented would have reduced rush changes, software workarounds and rollbacks, and risk of command errors.

Having mission-common visualization and resource tracking tools are important to provide a single frame of reference for all teams facilitating compatible solutions to end-to-end operations development from observation design through implementation.

Participation in meetings is necessary for all teams that desire to have a voice in decisions being made. Revisiting decisions can waste a lot of people's time.

## 6.2   Alice Logbook

For further Alice operational event details please refer to the Rosetta Alice Logbook that will be in the Rosetta Alice Archive. It contains details for Alice events, major spacecraft events, all observations by name and by date, and other information valuable to using and understanding Alice data.

# 7   Acknowledgements

Rosetta is an ESA mission with contributions from its member states and NASA. We would like to thank the members of the Rosetta Science Ground Segment and Mission Operations Center teams for their expert and dedicated help in planning and executing the Alice observations. We would like to thank the dedicated science and engineering staff at SwRI for all their contributions through the years. The Alice team acknowledges continuing support from NASA via Jet Propulsion Laboratory contract 1336850 to the Southwest Research Institute.

# 8 References

[1] S. A. Stern, D. C. Slater, J. Scherrer, J. Stone, M. Versteeg, M. F. A'Hearn, J. L. Bertaux, P. D. Feldman, M. C. Festou, Joel WM. Parker and O. H. W. Siegmund, "*ALICE*: THE *ROSETTA* ULTRAVIOLET IMAGING SPECTROGRAPH", Space Science Reviews (2007) 128, pp. 507–527

[2] Siegmund, O. W. H., Stock, J., Raffanti, R., Marsh, D., and Lampton, M., "UV and X-Ray Spectroscopy of Astrophysical and Laboratory Plasmas", Proceedings from the 10th International Colloquium, Berkeley, CA 3–5 February 1992, in Silver, E. and Kahn, S. (eds.), p. 383.

[3] Siegmund, O. W. H., Everman, E., Vallerga, J., and Lampton, M., "Optoelectronic Technologies for Remote Sensing from Space", Proceedings of the *SPIE*, SPIE, Bellingham, Washington, USA, 1987 Vol. 868, p. 17.

[4] Joel Wm. Parker, Andrew J. Steffl, S. Alan Stern. "ROSETTA-ALICE TO PLANETARY SCIENCE ARCHIVE INTERFACE CONTROL DOCUMENT" Rosetta PSA Document No. 8225-EAICD-01, March 2016 Rev 4 Chg 0

[5] Comet on 27 March 2016 – NavCam, Copyright ESA/Rosetta/NavCam – CC BY-SA IGO 3.0, Id 358019, Released 01/04/2016, http://www.esa.int/spaceinimages/Images/2016/04/Comet_on_27_March_2016_NavCam

[6] Comet's dusty environment, Copyright ESA/Rosetta/MPS for OSIRIS Team MPS/UPD/LAM/IAA/SSO/INTA/UPM/DASP/IDA, Id 345498, Released 13/08/2015, http://www.esa.int/spaceinimages/Images/2015/08/Comet_s_dusty_environment

[7] Mike Ashman, Maud Barthélémy, Laurence O'Rourke, Miguel Almeida, Nicolas Altobelli, Marc Costa Sitjà, Juan José García Beteta, Bernhard Geiger, Björn Grieger, David Heather, Raymond Hoofs, Michael Küppers, Patrick Martin, Richard Moissl, Claudio Múñoz Crego, Miguel Pérez-Ayúcar, Eduardo Sanchez Suarez, Matt Taylor, Claire Vallat, "Rosetta science operations in support of the Philae mission", Acta Astronautica, Volume 125, August–September 2016, Pages 41-64

[8] John Noonan, Eric Schindhelm, Joel Wm. Parker, Andrew Steffl, Michael Davis, S. Alan Stern, Zuni Levin, Sascha Kempf, Mihaly Horyani, "An investigation into potential causes of the anomalistic feature observed by the Rosetta Alice spectrograph around 67P/Churyumov–Gerasimenko", Acta Astronautica Volume 125, August–September 2016, Pages 3-10

[9] Andrea Accomazzo, Sylvain Lodiot, Vicente Companys, "Rosetta mission operations for landing", Acta Astronautica, Volume 125, August–September 2016, Pages 30-40

[10] Rosetta's journey around the comet, Copyright ESA, Released: 05/08/2016, http://www.esa.int/spaceinvideos/Videos/2016/08/Rosetta_s_journey_around_the_comet

[11] Brian A. Keeney, S. Alan Stern, Michael F. A'Hearn, Jean-Loup Bertaux, Lori M. Feaga, Paul D. Feldman, Richard A. Medina, Joel Wm. Parker, Jon P. Pineau, Eric Schindhelm, Andrew J. Steffl, M. Versteeg, and Harold A. Weaver, "H2O and O2 Absorption in the Coma of Comet 67P/Churyumov-Gerasimenko Measured by the Alice Far-Ultraviolet Spectrograph on Rosetta", 2017, Mon Not R Astron Soc stx1426. doi: 10.1093/mnras/stx1426

[12] Comet on 18 August 2014 – NavCam, ROS_CAM1_20140818T200718, Copyright ESA/Rosetta/NavCam – CC BY-SA IGO 3.0, horizontally inverted to match Alice histogram orientation, https://imagearchives.esac.esa.int/picture.php?/7204/categories/created-monthly-calendar-2014-8-18